\documentclass{aa}
\usepackage{graphicx}
\usepackage{txfonts}

\begin{document}
\title{Ages and metallicities of circumnuclear star formation regions
from Gemini IFU observations}

\author{O.~L. Dors Jr., T. Storchi-Bergmann, Rogemar A. Riffel, A.~A Schimdt} 

\offprints{Oli L. Dors Jr., \\
\email{oli.dors@ufrgs.br} }
\titlerunning{Circumnuclear star formation regions}

 \institute{Universidade Federal do Rio Grande do Sul, IF, CP 15051, 
 Porto Alegre 91501-970, RS, Brazil. \\}

\date{Received 29 octubre  2007 / Accepted 28 january 2008}
% \abstract{}{}{}{}{} 
% 5 {} token are mandatory   

\abstract
{}
{We derive the age and metallicity of circumnuclear star formation regions (CNSFRs) located
in the spiral galaxies \object{NGC\,6951} and \object{NGC\,1097},
and investigate the cause of the 
very low equivalent widths of emission lines
found for these regions.}
{We used optical two-dimensional spectroscopic data 
obtained with Gemini GMOS-IFUs and a grid of photoionization models
 to derive the the metallicities and ages of CNSFRs.}
{We find star formation rates in the range  0.002-0.14 $\rm M_{\odot}\: yr^{-1}$
and oxygen abundance of 12+log(O/H)$\approx$8.8 dex,   similar
to those  of most metal-rich nebulae located in the inner region of galactic disks.}
{We conclude that the very low emission-line equivalent widths
observed in CNSFRs are caused by the ``contamination'' of the continuum by (1) contribution of the 
underlying bulge continuum combined with (2) contribution from previous
episodes of star formation at the CNSFRs.}
\keywords{Galaxy: abundances --- Galaxies: individual: \object{NGC\,6951}, \object{NGC\,1097}
---ISM: abundances --- ISM: HII regions}
\maketitle
 
 %
%________________________________________________________________________

\section{Introduction}
%
%_____________

 The pioneer works by Morgan (\cite{morgan98}) and S\'ersic \& Pastoriza (\cite{sercic67})
have shown that many spiral galaxies host circumnuclear star formation regions (CNSFRs) and several
works have studied  these regions. 
 In one of the most recent works, Knapen (\cite{knapen05}) 
uses H$\alpha$ images of a sample of spiral galaxies
to study CNSFRs, as well as nuclear star formation regions. 
They found that CNSFRs are present in about 20$\%$ of spiral galaxies and almost always
occur within a barred host. In this paper we adopt
the definition  of  Knapen (\cite{knapen05}) to distinguish CNSFRs from nuclear
star formation regions: CNSFRs are the ones observed within circular areas with diameters 
of a few kpc around the galactic nucleus, while
nuclear star formation regions are observed in more internal areas a few hundred parsec across.
In particular, the formation of CNSFRs in barred galaxies seems to be  due to the 
 radial inflow of gas along bars to the galactic center (Roberts et al. \cite{roberts79};
 Friedli \& Benz \cite{friedli95}; Sakamoto et al. \cite{sakamoto99}; 
 Sheth et al. \cite{sheth05}; Jogge et al \cite{jogge05}). This gas is accumulated
in a ring structure between two inner Lindblad  resonances (Knapen et al. \cite{knapen95}).

Gas inflow is also necessary to feed the nuclear massive black hole 
 in active galactic nuclei (AGNs) and surrounding nuclear star formation regions, 
probably yielding the so-called AGN-starburst  connection
(Norman \& Scoville \cite{normam88}; Terlevich et al. \cite{terlevich90}; Heckman et al. \cite{heckman97};
Gonz\'alez Delgado et al. \cite{gonzalez98}; Ferrarese \& Merritt \cite{ferrarese00};
 Cid Fernandes et al. \cite{cid01};
Storchi-Bergmann \cite{thaisa01}; Heckman \cite{heckman04}).
Recent observational studies have shown evidence of gas inflow:
Storchi-Bergmann et al. (\cite{thaisa07}) and Fathi et al. (\cite{fathi06})
report the discovery of  gas streaming motions along spiral arms
towards  the LINER nuclei of the galaxies \object{NGC\,6951} and 
\object{NGC\,1097}, while Fathi et al. (\cite{fathi07}) found 
gas inflow to the nuclear star formation region of the galaxy \object{M\,83}.
Moreover, several works have found star formation around active 
nuclei (e.g. Riffel et al. \cite{riffel07}; Kauffmann et al. \cite{kauffmann03})   

In CNSFRs, the line ratio [\ion{O}{iii}]$\lambda$5007/H$\beta$
(hereafter [\ion{O}{iii}]/H$\beta$) is usually
weak or not observable and   the emission lines have lower equivalent widths than those of
disk \ion{H}{ii} regions (e.g. Diaz et al. \cite{diaz07};
Bresolin et al. \cite{bresolin99}; 
Bresolin \& Kennicutt \cite{bresolin97}; Kennicutt et al. \cite{kennicutt89}). 
The low value of [\ion{O}{iii}]/H$\beta$
has been attributed to  over-solar abundances
(e.g. Boer \& Schulz \cite{boer93}). At their location in 
the innermost parts of  galactic disks,
these objects are expected to be the most metal-rich star-forming regions. In fact, D\'{\i}az et al. (\cite{diaz07})
obtained long-slit observations for 12 CNSFRs and, using a semi-empirical method
to derivate of oxygen abundances, found that some of them 
have 12+log(O/H)=8.85, a value consistent with the 
maximum oxygen abundance value derived for central parts 
of spiral galaxies (Pilyugin et al. \cite{pilyugin07}).

Regarding the equivalent width of emission lines, Kennicutt et al. (\cite{kennicutt89})
found that the equivalent widths of H$\alpha$ EW(H$\alpha$) in CNSFRs are 
about 7 times lower than the ones
in disk \ion{H}{ii} regions. They proposed several scenarios to explain
this behavior, such as (i) deficiency
of high-mass stars in the initial mass function (IMF), 
(ii) a long timescale for star formation, and  (iii) high dust abundance.
However, neither of these mechanisms turned out to be the  dominant effect.
Unfortunately, spectroscopic data and abundance estimates  
are only available for a small number of CNSFRs.

With the goal of increasing the number of CNSFRs with determined gas 
abundances and  star formation rates, we have combined two-dimensional
integral field unit (IFU) data of  CNSFRs located in the galaxies 
\object{NGC\,1097} and \object{NGC\,6951} with a grid of photoionization models to derive 
their nebular gas properties. Using high-quality two-dimensional 
data, we could also investigate the cause of lower equivalent widths 
of emission lines observed in CNSFRs with respect to those of 
inner disk HII regions.
In Section~\ref{obs} we give details about the observational data we used,
in Section~\ref{phot} we present a description of the 
methodology used to derive the physical
parameters, while in Section~\ref{res} we present our results.
The discussion of the outcome and our final conclusions are
given in Sections~\ref{disc} and \ref{conc}, respectively.

\begin{figure}
\centering
\includegraphics[angle=-90,width=8cm]{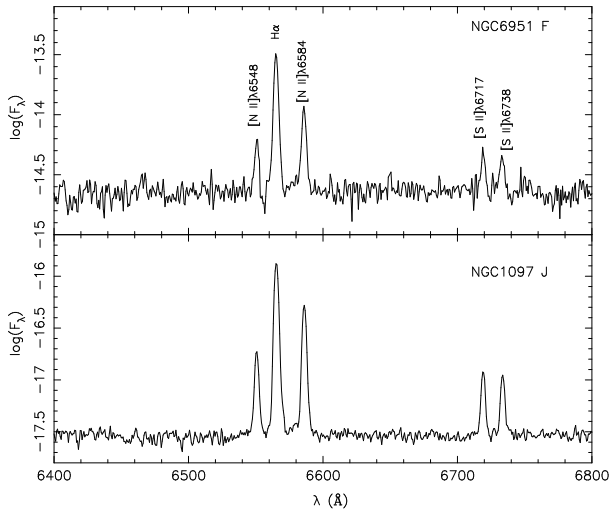}
\caption{Sample spectra: star-forming regions 
F in  \object{NGC\,6951} (top)  and J in \object{NGC\,1097} (bottom).}
\label{f1}
\end{figure}

\section{Observational data}
\label{obs}
 
The observational spectroscopic data  on \object{NGC\,1097} and \object{NGC\,6951} 
were drawn  from the works of Fathi et al. (\cite{fathi06}) and 
Storchi-Bergmann et al. (\cite{thaisa07}).
Basically these data were obtained
with IFUs of the Gemini Multi-Object Spectrographs (GMOS-IFU) on the Gemini 
South (for \object{NGC\,1097}) and North (for \object{NGC\,6951})
telescopes using the R400-G5325 grating
and r-G0326 filter, covering the spectral range 5600-7000 \AA \,
with a resolution of $\rm R \approx 2300$.
Three IFU fields were observed in each galaxy, covering an angular
field of 15\arcsec $\times$ 7\arcsec.
The reduction procedure
resulted in $50 \times 70$ spectra for each IFU field, each corresponding to an angular coverage of
0.1\arcsec $\times$ 0.1\arcsec. We adopted  distances for \object{NGC\,6951}
and \object{NGC\,1097}
of 24 Mpc and 17 Mpc (Tully \cite{tully88}), respectively, such that 1\arcsec corresponds  to
96 pc for the former and 68 pc for the latter galaxy.
The reader is referred to the papers above for a full description of the data. 

\begin{figure*}
\centering
\includegraphics[angle=-90,width=15cm]{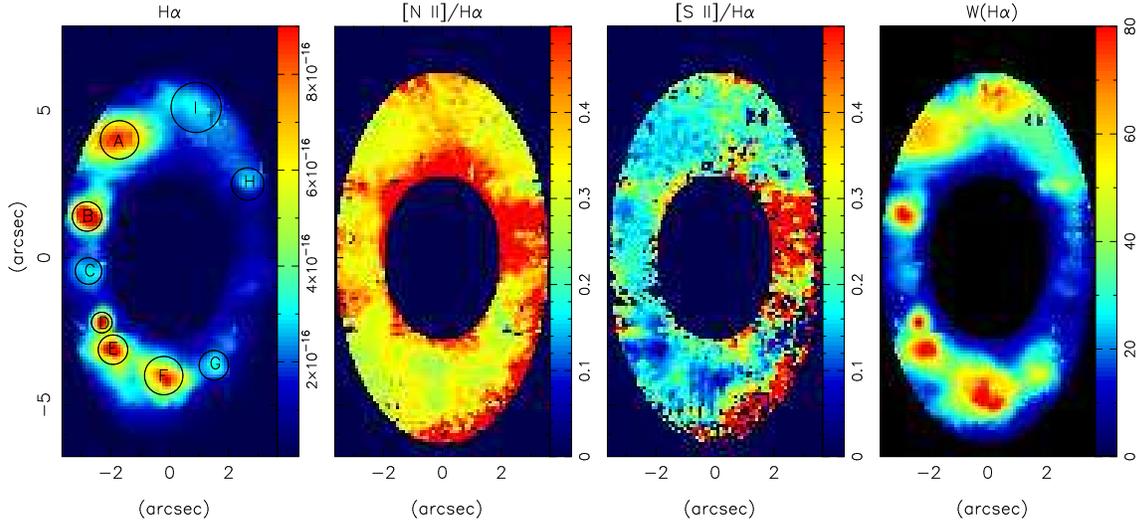}
\caption{Star-forming ring in \object{NGC\,6951}.
From left to right: map of the H$\alpha$ integrated flux ($\rm ergs \: cm^{-2} 
\: per \: pixel$); line ratio maps [\ion{N}{ii}]/H$\alpha$ 
[\ion{S}{ii}]/H$\alpha$ and EW(H$\alpha$). The circles and the letters identify
each region as defined in Section~\ref{obs}.}
\label{f2}
\end{figure*}

\begin{figure*}
\centering
\includegraphics[angle=-90,width=10cm]{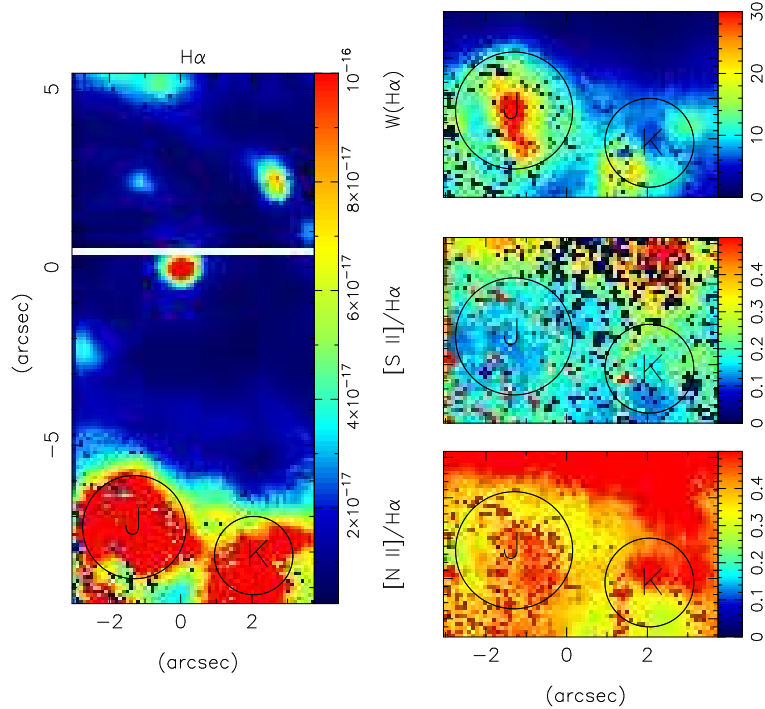}
\caption{{\it Left:} map of the H$\alpha$-integrated flux ($\rm ergs \: cm^{-2} 
\: per \: pixel$) showing the field covered by the IFU in \object{NGC\,1097},
which includes the nucleus and part of the ring (at the bottom). {\it Right:}
line ratio maps [\ion{N}{ii}]/H$\alpha$ 
[\ion{S}{ii}]/H$\alpha$ and EW(H$\alpha$) for the partial ring.
The circles and the letters identify
each region  defined in Section~\ref{obs}.} 
\label{f3}
\end{figure*}

 Figure~\ref{f1}  shows spectra of 
 CNSFRs  identified by the letters F and J  in Figs.~\ref{f2} and \ref{f3}
located in the circumnuclear rings of \object{NGC\,6951} and \object{NGC\,1097},
respectively. Using the observed spectra we built maps for flux distributions
in the  H$\alpha$ emission line and for the  line ratios [\ion{N}{ii}]$\lambda 6583$/H$\alpha$, 
([\ion{S}{ii}]$\lambda6717+\lambda6731$)/H$\alpha$, as well as for the EW(H$\alpha$).
 These maps are shown in Figs.~\ref{f2} and \ref{f3}.
Throughout the paper we identify the  line ratios above as [\ion{N}{ii}]/H$\alpha$ and 
[\ion{S}{ii}]/H$\alpha$.
For \object{NGC\,6951} the  observed
field completely covers  the circumnuclear ring.
The data of the ring were separated from those of the inner region
by fitting two ellipses to the H$\alpha$ image of \object{NGC\,6951}: one
to the inner border of the ring and the other to the outer border 
(see Storchi-Bergmann et al. \cite{thaisa07}). 
In the \object{NGC\,6951} H$\alpha$ map, we clearly see 
several regions of star formation.
We have identified nine of them as indicated in Fig.~\ref{f2}.
Since  for \object{NGC\,1097} the observation field covered only
part of the ring, we could only identify two  regions in this galaxy
(see Fig.~\ref{f3}).

For these measurements,
we assumed that each star formation region  has a circular symmetry, whose 
center corresponds to the peak in the H$\alpha$ intensity and the radius
defined as the one where the flux reaches 50 \% of the peak value.
Nevertheless, we can see from Fig.~\ref{f2} that the CNSFRs have  
irregular morphologies and the assumption above can result in some uncertainties
in the H$\alpha$ flux and in line ratio  measurements. 
To estimate how much flux we may be missing, for \object{NGC\,6951}, 
 we compared the sum of the
H$\alpha$ flux  of all regions (obtained from photometry within circular apertures) with 
the flux integrated over the  whole ring.  We found that the diffuse gas contribution
(from regions not covered by the circles) is 
 about 12 \%. A similar result was found  for individual \ion{H}{ii} regions 
by Oey \& Kennicutt (\cite{oey97}).
In relation to line ratios, we  measured the median values
considering different values for the radius of each region and
differences no greater than 10\% were found. We thus estimate that uncertainties
in both fluxes and line ratios due to uncertainties in the geometry of the
CNSFRs are about 10\%.
In Table \ref{tab1} we show the adopted area of each region, the luminosity of
H$\alpha$, and the median values of EW(H$\alpha$), [\ion{N}{ii}]/H$\alpha$, and 
[\ion{S}{ii}]/H$\alpha$.

\section{Photoionization models}
\label{phot}

To derive the metallicity, ionization parameter, and the
age  of the CNSFRs, 
we employed  the photoionization code  Cloudy/95.03 (Ferland
\cite{ferland02})  to build a grid of models covering a 
large space of nebular parameters. The models were built as
in Dors \& Copetti (\cite{dors06}),
with  metallicities $Z$=1.1, 1.0, 0.7, 0.6 
$Z_{\odot}$, and ionization parameter logarithm ($\log U$) of $-$2.0, $-$2.5, and $-$3.0  dex.
In each model a stellar cluster was assumed as the ionizing source with the  stellar energy distributions
obtained from the synthesis code $STARBURST99$ (Leitherer et al. \cite{leitherer99}).
We built models with stellar clusters  formed by  instantaneous and continuous bursts
with stellar upper mass limits of $M_{\rm up}$ = 20, 30 and 100 $M_{\odot}$, $Z$= 1.0 and 0.4
$Z_{\odot}$,
and ages ranging from 0.01 to 10 Myr with a step of 0.5 Myr.
Clusters older than about 10 Myr and formed instantaneously have 
no more O stars (Leitherer \& Heckman \cite{leitherer95}), so were not
considered in our models.
The metallicity of the nebula was matched to the closest available metallicity of the 
stellar cluster. The solar composition refers to
Asplund et al. (\cite{asplund05}) and corresponds to log(O/H) = $-$3.30.

\section{Results}
\label{res}

\begin{figure}
\centering
\includegraphics[angle=-90,width=8cm]{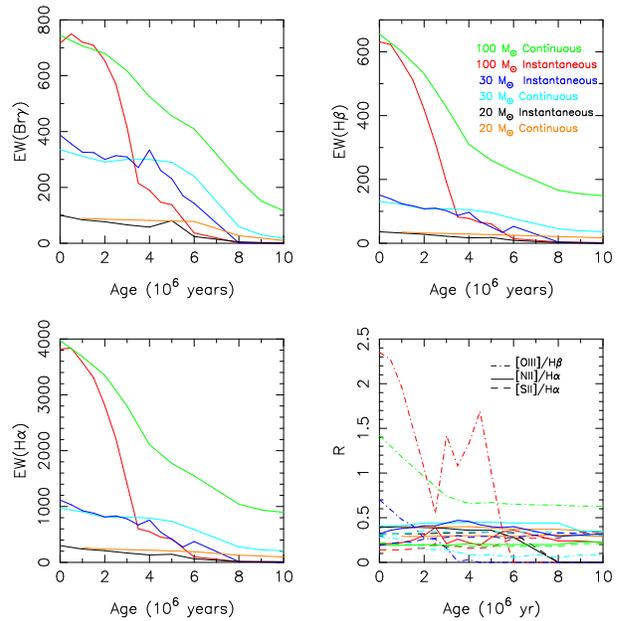}
\caption{Evolution of the EW(Br$\gamma$),  EW(H$\beta$), EW(H$\alpha$), 
  and relevant line ratios as predicted by our
 photoionization models considering different  $M_{\rm up}$ values
 and star formation regimes  as indicated, for $Z = Z_{\odot}$ and 
 $\log U =-2.5$.}
\label{f8}
\end{figure}

Since the observational data for \object{NGC\,6951} 
and \object{NGC\,1097} covers a restricted wavelength range, few
emission line ratios can be used in our analysis. 
Storchi-Bergmann et al. (\cite{thaisa94}) and Raimann et al. (\cite{raimann00})
showed that [\ion{N}{ii}]/H$\alpha$
can be used to obtain the oxygen abundance of star-forming galaxies.
However, this line ratio also dependents on the 
ionization parameter $U$ (Kewley \& Dopita \cite{kewley02}),
thus the combination with other line ratios is desirable in order to decrease the
uncertainties in the O/H estimates.  
Thus we used our photoionization  model grid  to build 
an [\ion{N}{ii}]/H$\alpha$ vs. [\ion{S}{ii}]/ H$\alpha$ diagram  to
derive $Z$ and $U$ of the CNSFRs located in \object{NGC\,6951} and \object{NGC\,1097}.
Recently, 
Viironen et al. (\cite{viironen07}) have shown that diagnostic diagrams 
using [\ion{N}{ii}]/H$\alpha$ and 
[\ion{S}{ii}]/ H$\alpha$ are a powerful tool for
estimating $Z$ and $U$ of star-forming regions.
Moreover, these line ratios are not dependent on the $M_{\rm up}$ and age assumed
in the models (see Fig.~\ref{f8}). 
 We point out that a cautionary note has been put forth by Mazzuca et al. (\cite{mazzuca06}) 
regarding the use of diagnostic diagrams including the line ratio [\ion{N}{ii}]/H$\alpha$:
they can yield degenerate values for the metallicity, as star-forming regions
with low $Z$ and $U$ have [\ion{N}{ii}]/H$\alpha$ 
values similar to regions with high $Z$ and $U$. However, CNSFRs
are known to have high abundances
(near solar) (see Sect.~\ref{metal}), ruling out the possibility
of low abundances. This is confirmed by the work of
Storchi-Bergmann et al. (\cite{thaisa96}), which shows an increasing
gas abundance towards the nucleus, including a few regions in  
circumnuclear rings.

\begin{figure}
\centering
\includegraphics[angle=-90,width=7cm]{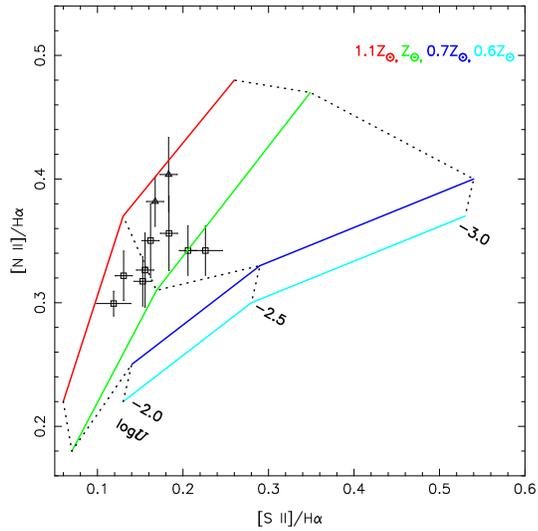}
\caption{Diagnostic diagram [\ion{N}{ii}]/H$\alpha$ vs. [\ion{S}{ii}]/H$\alpha$
showing the grid of photoionization models for an ionizing cluster with
age 2.5 Myr and $M_{\rm up} = 100 \: M_{\odot}$.
Solid lines connect curves of iso-$Z$, while dotted lines
connect curves of iso-$U$. The  values of $\log U$  and $Z$ are indicated.
Squares are the median values measured for the CNSFRs in \object{NGC\,6951} 
and triangles are those for  \object{NGC\,1097}.}
\label{f5}
\end{figure}

The results of our models compared to observational data for \object{NGC\,6951}
and \object{NGC\,1097} are shown in
Fig.~\ref{f5}, in which we considered  an instantaneous burst with 
age of 2.5 Myr and $M_{\rm up} = 100 \: M_{\odot}$.
 We can see that the observational data are 
well represented by models having
metallicities  $Z \approx Z_{\odot} $ and $\log U \approx -2.3$, 
 with uncertainties of about 0.03 and 0.1 dex, due to the
uncertainties in the line ratios discussed above.   
In Table~\ref{tab1} we show the  
$Z$ and $\log U$ values for each region, obtained by linear interpolation
from the model grid.
In this table we also present the values of electron density $N_{\rm e}$  computed 
from the observed line ratio [\ion{S}{ii}]$\lambda6717/\lambda6731$ and
adopting $T_{\rm e} = 10^{4}$ K (Osterbrock \cite{osterbrock89}).
The $N_{\rm e}$ values  cover  the range 50-600 $\rm cm^{-3}$, which is typical
 of \ion{H}{ii} regions (Copetti et al. \cite{copetti00}).

To obtain the ages of the CNSFRs, we used the 
observed values  of  equivalent widths  and  [\ion{O}{iii}]/H$\beta$, which
decrease  almost monotonically as a function of time 
(Magris et al. \cite{magris03}, Copetti et al. \cite{copetti86}).
In Fig.~\ref{f8} we plot the values predicted by the models 
for EW(Br$\gamma$), EW(H$\beta$), EW(H$\alpha$), as well as   
[\ion{N}{ii}]/H$\alpha$, [\ion{S}{ii}]/H$\alpha$ and [\ion{O}{iii}]/H$\beta$ 
versus  age. 
We considered instantaneous and continuous star formation bursts  with  
$M_{\rm up}$ = 20, 30 and 100 $M_{\odot}$, log~$U = -2.5$ and $Z= Z_{\odot}$. 
The values of the line ratios [\ion{N}{ii}]/H$\alpha$ and [\ion{S}{ii}]/H$\alpha$
were found to be approximately constant for the range of ages considered. However, 
[\ion{O}{iii}]/H$\beta$ shows a chaotic behavior between 2 and 6 Myr, mainly for 
$M_{\rm up}$=100 $M_{\odot}$ and instantaneous burst. 
This happens due to the increase in the number
of ionizing photons by the presence of Wolf Rayet stars 
in this   phase (Leitherer \& Heckman \cite{leitherer95}).
This behavior was also noted by Dopita et al. (\cite{dopita06}).
As no feature of Wolf Rayet stars was observed in our data and     
because [\ion{O}{iii}]/H$\beta$ is very dependent on many nebular 
parameters, it will not be
considered as an age tracer in this paper.
The use of the equivalent width of emission lines  seems to be more reliable for deriving ages
of star-forming regions,
since they basically depend on $M_{\rm up}$ and on the slope of the  IMF 
(Copetti et al. \cite{copetti86}).

Using the EW(H$\alpha$)  values shown in Table~\ref{tab1}, 
the CNSFR ages predicted by  instantaneous and continuous models   
are around 7 Myr and over 10 Myr, respectively,   independent
of the $M_{\rm up}$ assumed.  
Table~\ref{tab1} also presents  the values for 
the star formation rate (SFR) computed using the luminosity of the regions obtained
 in this paper and the  relation  
SFR$(M_{\odot}\:\rm year^{-1})$$= 7.9 \times \:L(\rm H\alpha)\:(ergs \:s^{-1})$ 
given by Kennicutt (\cite{kennicutt98}).

\begin{table*}
\centering
\centering
\caption{Physical properties of the CNSFRs in \object{NGC\,6951} and \object{NGC\,1097}}
\label{tab1}
\begin{tabular}{lccccccccc}
\hline
\noalign{\smallskip}       
Region    & $A$ ($\rm arcsec^{2}$) & $ L_{\rm H\alpha}$ ($10^{38}\:\rm erg \: s^{-1}$) &
$\rm [N\:II]/H\alpha$ &  $\rm [S\:II]/H\alpha$ &  SFR ($\rm M_{\odot}\: yr^{-1}$) &
EW(H$\alpha$) &   $Z/Z_{\odot}$  & $\log~U$    & $N_{\rm e}$($\rm cm^{-3}$) \\   
\noalign{\smallskip}  
\cline{1-10}
\noalign{\smallskip}
A & 3.0 &121 &0.32$\pm$0.03 &0.15$\pm$0.01 &0.09 &28 &1.04 &$-$2.3 &208$\pm$59 \\
B & 1.5&68.2 &0.35$\pm$0.03 &0.16$\pm$0.01 &0.05 &31 &1.05 &$-$2.2 &287$\pm$151 \\
C & 1.4&23.1 &0.35$\pm$0.03 &0.18$\pm$0.01 &0.02 &16 &1.04 &$-$2.4 &297$\pm$134 \\
D & 0.6&25.0 &0.31$\pm$0.02 &0.15$\pm$0.01 &0.02 &39 &1.03 &$-$2.3 &525$\pm$103 \\
E & 1.4&59.7 &0.32$\pm$0.02 &0.13$\pm$0.01 &0.05 &45 &1.05 &$-$2.1 &354$\pm$24 \\
F & 2.9&94.5 &0.29$\pm$0.01 &0.11$\pm$0.02 &0.07 &26 &1.04 &$-$2.0 &148$\pm$40  \\
G & 0.9&16.5 &0.31$\pm$0.01 &---            &0.01 &32 &---  &---    &273$\pm$36   \\
H & 0.8&12.2 &0.34$\pm$0.02 &0.22$\pm$0.02 &0.01 &31 &0.94 &$-$2.6 &68$\pm$10  \\
I & 3.0&47.6 &0.34$\pm$0.02 &0.20$\pm$0.01 &0.04 &38 &0.99 &$-$2.6 &138$\pm$29  \\
J & 8.9&45.8 &0.38$\pm$0.02 &0.16$\pm$0.01 &0.03 &12 &1.08 &$-$2.1 &292$\pm$45  \\
K & 1.8&2.06 &0.40$\pm$0.03 &0.18$\pm$0.01 &0.002&7  &1.09 &$-$2.2 &223$\pm$149  \\
\noalign{\smallskip}
\hline 
\end{tabular}
\end{table*}

\section{Discussion}
\label{disc}

\subsection{Metallicities}
\label{metal}

Acording to our  models, the  CNSFRs of \object{NGC\,6951} and \object{NGC\,1097} have
solar or slightly higher  solar metallicities.
Considering an uncertainty on the order of 0.1 dex due to  
 the method used to estimate  abundances
(see Evans \cite{evans86}, Dors \& Copetti \cite{dors05}, \cite{dors06}), 
our abundance values are comparable to 
the maximum oxygen abundance derived for central parts of spiral galaxies 
of 12+log(O/H) $\approx$ 8.87 (Pilyugin et al. \cite{pilyugin07}).
 Similarly, Mazzuca et al. (\cite{mazzuca06}) present an emission-line diagnostic analysis
of the nuclear starburst ring of \object{NGC\,7742} and find
roughly solar abundances.

Recently,  Sarzi et al. (\cite{sarzi07}) presented a study  
of CNSFRs located in eight galaxies
which  include the regions B, E and F of our sample.
They also used   empirical methods to derive the metallicities 
of these regions and find abundances similar to the
ones derived in this paper. D\'{\i}az et al. (\cite{diaz07}) 
use a semi-empirical abundance calibration to 
find that the CNSFRs in \object{NGC\,2903}, \object{NGC\,3351}, and
\object{NGC\,3504} have metallicities comparable to the ones derived 
in this paper.  
Until now, only a case of very low abundance 
on the order of 0.2-0.4 $Z_{\odot}$ in the CNSFRs
located in \object{NGC\,3310} was reported by Pastoriza et al. (\cite{pastoriza93}). 
This low abundance can be understood as due to the fact that this galaxy has
an interaction history (Elmegreen et al. \cite{elmegreen02}), 
through which neutral gas may have been accreted to the ring. 

\begin{figure}
\centering
\includegraphics[angle=-90.0,width=7cm]{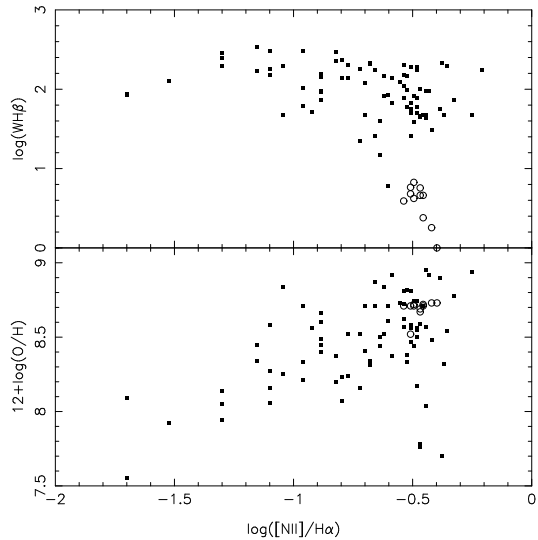}
\caption{Oxygen abundance vs. [\ion{N}{ii}]/H$\alpha$  (lower panel) and
EW(H$\beta$) vs. [\ion{N}{ii}]/H$\alpha$ (upper panel). The squares represent
data collected from the literature, while the circles
represent the data of the CNSFRs analyzed in this paper.}
\label{f9}
\end{figure}

Due to the fact that CNSFRs have very low excitation, 
with [\ion{O}{iii}]/H$\beta < 1$ (e.g. H\"agale et al. \cite{hagele07}, 
P\'erez et al. \cite{perez00}) and  to their central location 
in the galactic disk, it is expected that these objects have somewhat higher abundance than
is derived for inner disk \ion{H}{ii} regions.
To check this, we have plotted the [\ion{N}{ii}]/H$\alpha$ in Fig.~\ref{f9} (lower panel)
against oxygen abundance
for some  disk \ion{H}{ii} regions located in 13 spiral galaxies.
This sample includes the most metal-rich nebulae in which  temperature sensitive emission lines
have been detected. 
The data were taken from Bresolin (\cite{bresolin07}), Bresolin et al. (\cite{bresolin05}, \cite{bresolin04}),
Kennicutt et al. (\cite{kennicutt03}), Castellanos et al. (\cite{castelanos02}),
and Skillman et al. (\cite{skillman96}). With the exception of the oxygen abundances
from Skillman et al. (\cite{skillman96}), which were recalculated  by us using 
the $P$-method (Pilyugin \cite{pilyugin01}), all O/H values were obtained
using direct derivation of the electron temperature. Also included in Fig.~\ref{f9}
are the data of our sample. 
We can see that the CNSFRs show the highest oxygen abundances, together
with some disk nebulae, a result also obtained by D\'{\i}az et al. (\cite{diaz07}). 
Since abundance determinations 
via photoionization models are overestimated by a factor of 0.1-0.4 dex
when compared to direct determinations (Dors \& Copetti \cite{dors05}),
our abundance estimates should be interpreted as upper limits.
 If CNSFRs are included in the computation of abundance gradients in spirals,
we would expect therefore a plateau at very small galactocentric distances.

\subsection{Star formation history}
\label{sf}

The H$\alpha$ luminosity values found  for the CNSFRs in this paper are 
in the range $2 \times 10^{38}- 1.2 \times 10^{40} \: \rm erg \: s^{-1}$, compatible with values 
of regions in circumnuclear rings of other spiral galaxies (e.g. D\'{\i}az et al. \cite{diaz07}).
These luminosity values yield star formation rates of  
0.002-0.1 $\rm M_{\odot}\: yr^{-1}$, characterizing a  moderate
star-forming regime. Averaging the SFR values in Table~\ref{tab1}, we
find 0.04 $\rm M_{\odot}\: yr^{-1}$, a similar
value to the one found by Ho et al. (\cite{ho97}) for a sample
of nuclear star-forming regions and by Shi et al. (\cite{shi06}),
who studied a sample of 385 circumnuclear star-forming regions in galaxies
with different Hubble types. 

We cannot conclude from our data alone whether the star
formation has been continuous or instantaneous in the rings, although the former
scenario has not been favored by previous studies.
For example, Sarzi et al. (\cite{sarzi07}) use absorption-line index diagrams
to show that the hypothesis of a constant star formation in their sample 
can be ruled out. 
 Allard et al. (\cite{allard06}) find
that the star formation in the ring of \object{M\,100} 
have occurred in a series of short bursts for the past 500 Myr or so.
These results can be understood within the following scenario.
Once massive stars are formed, they cause turbulence and
heating in the molecular cloud, inhibiting further star formation
(Hartmann et al. \cite{hartmann01}, Blitz \& Shu \cite{blitz80}).
Since no systematic age sequence is observed along the \object{NGC\,6951}
ring, the star formation mode seems to be as predicted by the {\it popcorn model}
(B\"oker et al. \cite{boker07}), in which the entire 
ring begins to form stars at about the same time or at random times.
 A similar result was also derived by Mazzuca et al. (\cite{mazzuca07}),
who used photometric H$\alpha$ imaging of 22 nuclear rings
 to find that only a very small fraction of them   
show age sequences along the ring.

\subsection{Age estimates}
\label{age}
 
Our  modeling gives ages of 
about 7.0 Myr for an instantaneous burst and more than 10 Myr
for a continuous burst, with the CNSFRs in \object{NGC\,1097}  slightly
older than  in \object{NGC\,6951}. To compare  our age estimates
with those 
of other CNSFRs, we collected equivalent widths of more CNSFRs from the
literature  and estimated their
ages using Fig.~\ref{f8}. The data were extracted from 
D\'{\i}az et al. (\cite{diaz07}) for \object{NGC\,2903} and \object{NGC\,3505},
H\"agele et al. (\cite{hagele07}) for \object{NGC\,3351},
Allard et al. (\cite{allard06}) for \object{M\,100},
Reunanen et al. (\cite{reunanen00}) for \object{NGC\,7771},
and  Wakamatsu \& Nishida (\cite{wakamatsu}) for \object{NGC\,4314},
which give EW(H$\beta$), EW(H$\alpha$), and EW(Br$\gamma$)
in the range 1-17, 10-40, and 5-60 \AA, respectively. 
The predicted ages
for these EW values are of 5-8 Myr and more than 10 Myr 
for instantaneous and continuous bursts, respectively,
thus  similar to the ones we derived for the CNSFRs in the present paper.
 Somewhat higher EW values have been obtained by Mazzuca et al. (\cite{mazzuca07})
for HII regions in nuclear rings of more than 20 galaxies.
On the other hand, studies also utilizing photoionization models
give ages  of 1.5-3 Myr for \ion{H}{ii} regions located in 
galactic disks  and  for H\,II galaxies 
(Dors \& Copetti \cite{dors06}; Stasinska \& Izotov
\cite{stasinska03}; Bresolin et al. \cite{bresolin99}; 
Copetti et al. \cite{copetti85}).  
We have also compared the ages of our sample
with those of disk  nebulae of similar metallicity. 
In Fig.~\ref{f9} we show log(EW(H$\beta$))
and 12+log(O/H) plotted against log([\ion{N}{ii}]/H$\alpha$)
for these disk \ion{H}{ii} regions, together with our data 
for the CNSFRs.
We converted the EW(H$\alpha$)  values
shown in Table~\ref{tab1}  in EW(H$\beta$) using the 
relation EW(H$\beta$)$\approx 0.15$ EW(H$\alpha$)
obtained from our models.  
We can see that the EW(H$\beta$) of the CNSFRs are lower
by a factor of about 30 than those of their counterpart disk \ion{H}{ii} regions.
Assuming a EW(H$\beta$)= 100 \AA \, for the innermost disk regions and using
Fig.~\ref{f8}, we find that these regions are  about 3 Myr younger than 
the CNSFRs.

\subsection{The low EW values}
\label{age}

The age differences found above may not be real, since  
the very low equivalent widths in CNSFRs stem from other effects  
than the age. Kennicutt et al. (\cite{kennicutt89}) pointed out 
 possible mechanisms such as (i) deficiency
of high-mass stars in the initial mass function, (ii) very high dust abundance,
(iii) contamination of the continuum by contribution from the bulge or 
other underlying  stellar populations, and  (iv) continuous star formation.
These hypotheses are discussed below. 

Regarding the hypothesis of the deficiency
of high-mass stars, since  $M_{\rm up}$ of an ionizing cluster is dependent
on the metallicity
(Larson \& Starrfield \cite{larson71}; Kahn \cite{kahn74}; 
Shields \& Tinsley \cite{shields76}; Stasi\'nska
\cite{stasinska80}; Vilchez \& Pagel \cite{vilchez88}; Campbell \cite{campbell88};  
Bresolin et al. \cite{bresolin99};
Dors \& Copetti \cite{dors03}, \cite{dors05}, \cite{dors06}), 
CNSFRs and innermost disk nebulae should have about same $M_{\rm up}$. 
It is unlikely  that the discrepancy in EWs is only due to variation
in $M_{\rm up}$. Even assuming that the ionizing cluster of 
disk nebulae have $M_{\rm up}$ = 100 $M_{\odot}$ and the ones of
CNSFRs have $M_{\rm up}$ = 20 $M_{\odot}$, the age difference
remains. Also a variation in the IMF can be discarded, since
stellar clusters appear to form with a universal IMF slope (Kroupa
\cite{kroupa07}, \cite{kroupa02}, Chabrier \cite{chabrier03}).

The hypothesis of very high dust abundance in CNSFRs is not favored because they have 
similar Balmer decrements to those observed for disk \ion{H}{ii} regions,
suggesting that this is not the main source of the low EWs in CNSFRs
(Kennicutt et al. \cite{kennicutt89}).

Concerning the contamination of the continuum,
several works have found the presence
of underlying stellar populations from previous generations of
stars in CNSFRs, which can decrease
the equivalent widths and yield no real age estimates
from Fig.~\ref{f8}.
For example, Buta et al. (\cite{buta00}) found a spread in ages of  about 5 to 200 Myr for 
clusters in the circumnuclear ring of \object{NGC\,1326}, with the largest number of clusters 
younger than 20 Myr. 
In the case of  \object{NGC\,6951} and with measurements of the equivalent width
of calcium triplet at $\approx$8500 \AA,
P\'erez et al. (\cite{perez00}) found evidence of a 
population of red supergiant stars
with ages of 10 to 20 Myr in the ring.

An underlying stellar population can contribute in two ways to the low EWs
by causing a Balmer absorption and  yielding an extra contribution to the continuum.
To estimate the effect of an underlying Balmer absortion, we 
 constructed models of stellar populations (using the {\it Padova 1994 tracks})
 with ages in the range 10-1000 Myr. We normalized the continuum of
 each model to the observed one and subtracted them from the observed spectra
 of the CNSFRs of \object{NGC\,6951}. For all cases, the increase in the EW values
due to subtracting of an underlying absorption
was at most 10\%, much less than needed to
compensate for the low observed EW in CNSFRs.

Determining of the continuum contribution from underlying stellar
populations is not an easy task, 
since two types of stellar populations can be contributing:
the stellar population from the bulge and stars
formed in previous episodes of star formation in the ring.
To evaluate the contribution of the underlying bulge,
 we performed aperture photometry
on the continuum images (not shown) measuring both the CNSFR continuum and that
of the bulge in the surrounding regions. 
We concluded that the bulge
contributes with $\approx$ 75 \% of the total flux in the case of \object{NGC\,6951}
 and with 50 \% in \object{NGC\,1097}.
The corresponding corrections  are an increase by a factor of 2
and 4 in EW(H$\alpha$) for \object{NGC\,1097}  
and \object{NGC\,6951}, respectively. 
These factors increase the EW(H$\alpha$) in \object{NGC\,6951}
from average values of $\approx$30 \AA \, to $\approx$120 \AA \, 
and from $\approx$10 \AA \, to $\approx$20 \AA \, in  
\object{NGC\,1097}. The corresponding changes in age are from
$\approx$7 Myr to $\approx$6 Myr in \object{NGC\,6951} and no change
for \object{NGC\,1097} ($\approx$7 Myr). Thus the contribution from
the bulge to the continuum also does not explain   the low
EW's in CNSFRs.
Stauffer (\cite{stauffer81}) also studied
the bulge contamination using  aperture photometry 
of galaxy nuclei to measure the background contribution 
to the continuum flux, and
Kennicutt et al. (\cite{kennicutt89})
used those results to find a correction factor similar to
the one found by us.

\begin{figure}
\centering
\includegraphics[angle=-90.0,width=7cm]{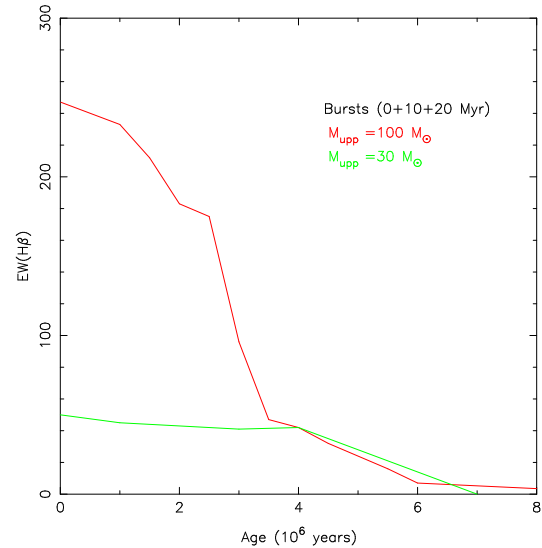}
\caption{Evolution of the EW(H$\beta$) as predicted by 
 photoionization models considering 
three  episodes of star formation as indicated. 
The ages correspond to the youngest burst.}
\label{f10}
\end{figure}

Several generations of stars may coexist in CNSFRs   because 
the tidal effects are not strong enough  to disrupt the cluster
at the radius of the ring, allowing several generations of stars
 to coexist (Buta et al. \cite{buta00}). Thus the continuum may have
a contribution from previous bursts of star formation in the rings.
To investigate the effect of this continuum, we
 built photoionization models to represent a scenario
where three bursts of star formation have occurred, with ages 
of 0.01, 10, and 20 Myr for two values of $M_{\rm up}$: 100 and 30 $M_{\odot}$.
This is illustrated in  Fig.~\ref{f10}, which  shows
the evolution of EW(H$\beta$) as a function of time. It can be
seen that the EW values decrease considerably when compared with those
of a single burst (Fig.~\ref{f8}).
If we consider a typical EW(H$\beta$) for CNSFRs of 10 \AA \, (not corrected for
 the   underlying bulge contribution)
and compute the age from Fig.~\ref{f10},
we find ages of 6-7 Myr, similar to the values derived above.

We conclude  that none of the mechanisms discussed above   
alone yield   the low EW values (and resulting larger age) observed in CNSFRs
when compared to disk \ion{H}{ii} regions. 
However, it could be that a combination of more than one mechanism 
can explain the lower EWs in CNSFRs. For example, we can consider the 
effects of both the dilution by the bulge and by previous bursts of star formation.
A typical EW(H$\beta$)= 10 \AA \, in \object{NGC\,6951} will increase to 40 \AA \, if we correct
for contamination by the bulge. Then, we can consider the effect of previous generation
of stars in the ring using Fig.~\ref{f10} instead of Fig.~\ref{f8} to obtain the age. 
We obtain an age of 3-4 Myr, a value  in consonance with ages of the 
innermost disk regions. In this way we 
eliminate the apparent difference in age between CNSFRs and inner-disk
 \ion{H}{ii} regions.

\section{Conclusions}
\label{conc}

We have used optical IFU spectroscopic data  of CNSFRs in
 \object{NGC\,6951} and \object{NGC\,1097},
which  compared to a grid of photoionization models, 
in order to obtain physical parameters of these regions. 
We find that the star formation rates 
are in the range  0.002-0.14 $\rm M_{\odot}\: yr^{-1}$, thus 
covering a range from very low to moderate
star-forming regime. We also find that the CNSFRs have
oxygen abundances of 12+log(O/H)$\approx$8.8,  similar
to those of the most metal-rich nebulae  located in inner parts 
the disks of spiral galaxies.

We investigated the cause of the very low  equivalent width
of emission lines found in CNSFRs, which suggests an older age for
these objects than for disk \ion{H}{ii} regions. 
 Among the several scenarios invoked to explain 
this fact, we have concluded that the contamination of
the continua of CNSFRs by underlying contributions from both
old bulge stars and stars formed in the ring in previous 
episodes of star formation (10-20 Myr)
yield the observed low  equivalent widths. Correcting for these contributions,
there are not any more significant differences in ages between the CNSFRs
and inner disk \ion{H}{ii} regions.

 \begin{acknowledgements}
 We thank the anonymous referee for valuable suggestions which helped to 
improve the paper. This work was based on observations obtained at the Gemini Observatory, which is operated by the
Association of Universities for Research in Astronomy, Inc., under a cooperative agreement
with the NSF on behalf of the Gemini partnership: the National Science Foundation (United
States), the Science and Technology Facilities Council (United Kingdom), the
National Research Council (Canada), CONICYT (Chile), the Australian Research Council
(Australia), CNPq (Brazil) and SECYT (Argentina)
\end{acknowledgements} 
 
 {}

\end{document}